\documentclass[aip, amsmath, amssymb, reprint]{revtex4-2}
\usepackage{graphicx}
\usepackage{soul, xcolor}
\usepackage[normalem]{ulem}
\usepackage{cancel}
\usepackage{dcolumn}
\usepackage{bm}
\usepackage[utf8]{inputenc}
\usepackage[T1]{fontenc}
\usepackage{mathptmx}
\graphicspath{{./}}
\makeatletter
\let\@fnsymbol\@fnsymbol@latex
\@booleanfalse\altaffilletter@sw
\makeatother
\begin{document}
	
	\title{Frequency-tunable Kerr-free three-wave mixing with a gradiometric SNAIL}
	\author{A. Miano}
	\email[Author to whom correspondence should be addressed:]{sandro.miano@yale.edu}
	\affiliation{Department of Applied Physics, Yale University, New Haven, Connecticut 06520, USA}
	\author{G. Liu}
	\affiliation{Department of Applied Physics, Yale University, New Haven, Connecticut 06520, USA}
	\author{V. V. Sivak}
	\affiliation{Department of Applied Physics, Yale University, New Haven, Connecticut 06520, USA}
	\author{N. E. Frattini}
	\affiliation{Department of Applied Physics, Yale University, New Haven, Connecticut 06520, USA}
	\author{V. R. Joshi}
	\affiliation{Department of Applied Physics, Yale University, New Haven, Connecticut 06520, USA}
	\author{W. Dai}
	\affiliation{Department of Applied Physics, Yale University, New Haven, Connecticut 06520, USA}
	\author{L. Frunzio}
	\affiliation{Department of Applied Physics, Yale University, New Haven, Connecticut 06520, USA}
	\author{M. H. Devoret}
	\affiliation{Department of Applied Physics, Yale University, New Haven, Connecticut 06520, USA}	
	\begin{abstract}
		Three-wave mixing is a key process in superconducting quantum information processing, being involved in quantum-limited amplification and parametric coupling between superconducting cavities. These operations can be implemented by SNAIL-based devices that present a Kerr-free flux-bias point where unwanted parasitic effects such as Stark shift are suppressed. However, with a single flux-bias parameter, these circuits can only host one Kerr-free point, limiting the range of their applications. In this Letter, we demonstrate how to overcome this constraint {by introducing the} gradiometric SNAIL, a doubly-flux biased superconducting circuit {in} which both effective inductance and Kerr coefficient can be independently tuned. Experimental data show the capability of the gradiometric SNAIL to suppress Kerr effect in a three-wave mixing parametric amplifier over a continuum of flux bias points corresponding to a 1.7 GHz range of operating frequencies.
	\end{abstract}
	\maketitle
	Superconducting circuits for quantum computation rely on Josephson Tunnel Junctions (JTJs) to implement nonlinear operations. For microwave signals at frequencies significantly lower than its plasma frequency, a JTJ acts as a nonlinear inductance whose response can be tuned \emph{in situ} when a proper phase bias is applied across it. This is achieved by threading a magnetic flux through a superconducting loop involving the JTJ. More generally, a one-loop JTJ circuit with two leads attached constitutes a dipole that can be characterized by a tunable potential energy 	
	\begin{equation} \label{dipole_potential}
		U(\bm{\varphi}) = \sum_{n=2}^\infty \frac{c_n(\Phi)}{n!}(\bm{\varphi}-\bm{\varphi}_0)^n
	\end{equation}
	where $\bm{\varphi}$ is the gauge-invariant phase drop across the dipole, $\Phi$ is the magnetic flux threaded through the loop, $c_n(\Phi) = \left.d^nU/d\bm{\varphi}^n\right|_{\bm{\varphi}_0}$ are Taylor expansion coefficients, and $\bm{\varphi}_0$ is the value of $\bm{\varphi}$ that minimizes the potential energy.
	In expression \eqref{dipole_potential}, the coefficients $c_n(\Phi)$ can be varied with flux in order to select a particular combination of linear and nonlinear effects. For instance, a superconducting dipole can be configured to offer a certain value of inductance $\propto 1/c_2(\Phi)$, three-photon processes ${\propto}c_3(\Phi)$ or four-photon processes ${\propto}c_4(\Phi)$ efficiencies. Such tunable dipoles have been implemented with rf-SQUIDs \cite{zorin_traveling-wave_2017} and with SNAILs (Superconducting Nonlinear Asymmetric Inductive eLement) \cite{frattini_3-wave_2017} employed for three-wave \cite{frattini_optimizing_2018,sivak_kerr-free_2019,sivak_josephson_2020} and four-wave mixing \cite{ranadive_reversed_2021} operations. 
	
	The SNAIL, represented on the left of Fig. 1\textbf{a}, is realized by shunting an array of typically three indentical JTJs, each one with Josephson energy $E_\mathrm{J}$, with a single JTJ with Josephson energy $\alpha E_\mathrm{J}$, with $\alpha<1/3$. This constraint is required {to ensure} that the potential energy \eqref{dipole_potential} has a single minimum in the interval $[-3\pi,3\pi]$ for every value of external flux $\Phi$.
	The application of a flux-bias to a SNAIL makes it a versatile superconducting dipole capable of implementing different combinations of nonlinear effects: three-wave mixing amplification with suppressed Kerr constant $K(c_2,c_3,c_4)$ \cite{frattini_optimizing_2018, sivak_kerr-free_2019}, a combination of three-wave mixing and Kerr \cite{kerrcat} or four-wave-mixing with reversed Kerr {constant} \cite{ranadive_reversed_2021}. However, once a flux-bias point is selected, for instance, to obtain a desired value of the Kerr constant, the values of $c_2$ and $c_3$ will be fixed as a consequence, restricting the range of possible applications of a SNAIL-based circuit.
	Is it possible to decouple the flux dependencies of two parametric processes and, for instance, achieve three-wave mixing amplification with suppressed Kerr constant at any desired value of {resonance frequency}?
	
	In this Letter, we show how to implement this type of tunability with a doubly-flux biased Josephson device, namely the Gradiometric SNAIL (G-SNAIL).	
	{Strongly} influenced by two recent superconducting circuits biased by two \emph{in situ} magnetic fluxes \cite{miano_symmetric_2019, lescanne_exponential_2020}, our device is capable of selecting a combination of two different parametric processes thanks to an additional tuning knob with respect to single-flux biased devices.
	For example, a G-SNAIL can implement the same linear response $c_2$ for different combinations of bias fluxes, corresponding to different combinations of $c_3$ and $c_4$. {In particular,} a G-SNAIL could be capable, for instance, of keeping the Kerr constant $K(c_2,c_3,c_4)$ suppressed while varying $c_2$.
	{We want to clarify that the \emph{symmetric rf-SQUID}\cite{miano_symmetric_2019} first demonstrated the possibility to doubly-flux bias a Josephson device for 3WM applications at microwave frequencies. However, it was operated as the building block of a TWPA and lacked a quantitative demonstration of independent tuning of two parametric processes.
		With the G-SNAIL we propose a minimal and more compact realization of a similar idea in a resonant parametric amplifier.}
	
	A G-SNAIL is obtained
	{by shunting a SNAIL with another small junction with Josephson energy $\alpha E_\mathrm{J}$,}
	forming the circuit on the right of Fig. 1\textbf{a}. This produces a dipole with a central array of three JTJs with energy $E_\mathrm{J} = \Phi_0I_\mathrm{c}/2\pi$ (purple junctions in Fig. 1\textbf{a}), shunted by two JTJs with energy $\alpha E_\mathrm{J}$ (blue junctions in Fig. 1\textbf{a}). Here, $\Phi_0$ is the magnetic flux quantum and $I_\mathrm{c}$ is the critical current of a JTJ in the central array of the G-SNAIL. The two loops formed by this configuration can then be independently flux biased.
	{This configuration is equivalent to a symmetric dc-SQUID shunted by an array of three JTJs. However, a G-SNAIL with a symmetric layout can be more efficiently and independently flux-biased by two symmetrically placed on-chip current lines, as discussed later in the manuscript.}
	
	{To derive the potential energy for a G-SNAIL, we choose the convention in which the phase variable $\varphi$ is equal to the phase drop across the central array (purple junctions in Fig. 1\textbf{a}). Consequently, the magnetic fluxes $\Phi_{1(2)}$, respectively threaded to the left and right loops of the G-SNAIL, will only appear in the contributions of the two small junctions to the total potential energy. After applying standard trigonometric transformations to separate the $\cos\varphi$ and $\sin\varphi$ contributions arising from the flux-tunable terms, the G-SNAIL potential energy can be expressed as}
	\begin{equation}\label{potential}
		\frac{U(\bm{\varphi})}{E_\mathrm{J}} =-3\cos{\frac{\bm{\varphi}}{3}}-\left[b_\mathrm{e}(\varphi_1,\varphi_2)\cos\bm{\varphi}+b_\mathrm{o}(\varphi_1,\varphi_2)\sin\bm{\varphi}\right]
	\end{equation}
	\begin{figure}
		
		\label{fig1}
		\includegraphics[width = \columnwidth]{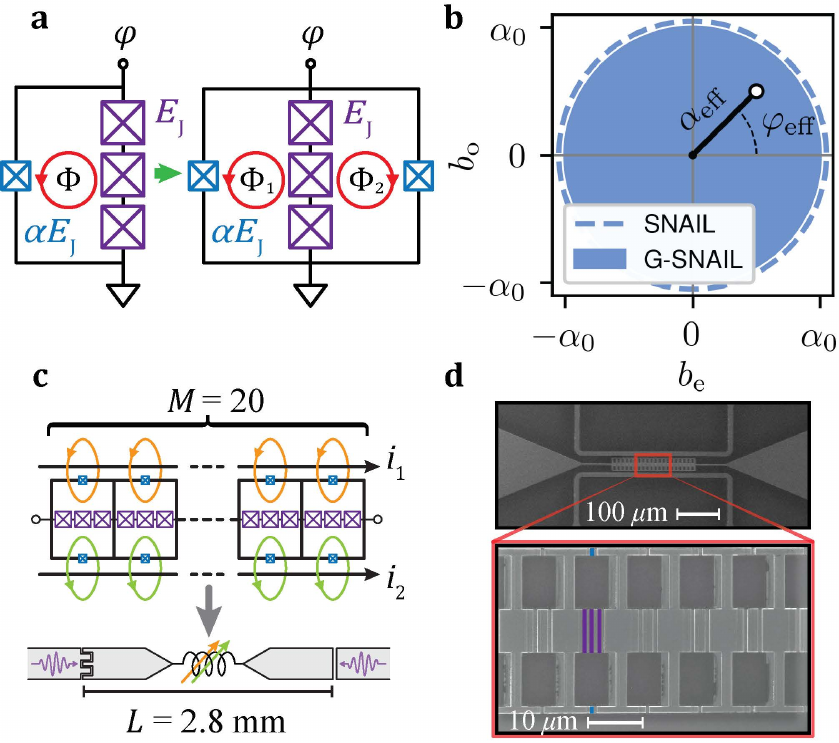}
		\caption{(a) A SNAIL (on the left) is transformed into a G-SNAIL (on the right) by
		{shunting a SNAIL with another blue junction.} 
		A G-SNAIL (on the right) has a central array of three JTJs, each one with energy $E_\mathrm{J}$, shunted on each side with a smaller junction of energy $\alpha E_\mathrm{J}$. The obtained loops are biased with two external fluxes $\Phi_{1(2)}$. (b) Bias parameters plane: tunable with $\varphi_{\pm}$ provided by external magnetic fluxes, $b_\mathrm{e}$ and $b_\mathrm{o}$ can be chosen inside the whole unit circle for a G-SNAIL while, for a SNAIL, the same parameters are constrained to the circle perimeter.
			(c) Gradiometric SNAIL parametric amplifier: an array of $M=20$ G-SNAILs is flux biased with two on-chip current lines and embedded in the center of a $L = \lambda/2$ resonator. Resonator is strongly coupled to a probe port (left) and weakly coupled to a pump port (right). (d) SEM image of the device: an array of 20 G-SNAILs fabricated with Dolan bridge technique. Three junctions with Josephson energy $E_\mathrm{J}$ (purple) form the central array and and shunted with a junction with energy $\alpha E_\mathrm{J}$ (blue) on each side.}
	\end{figure}
	where $\varphi_{1(2)} = 2\pi\Phi_{1(2)}/\Phi_0$
	{and $b_\mathrm{e(o)}$ are flux-tunable bias parameters which weight even and odd order nonlinear terms.}
	Introducting $\varphi_\pm = (\varphi_1 \pm \varphi_2)/2$,
	{we can define} the effective quantities
	\begin{equation}
		\label{eff_quantities}
		\begin{split}
			\alpha_\mathrm{eff} &\stackrel{\text{def}}{=} {\alpha_0}|\cos{\varphi_-}|\\
			\varphi_\mathrm{eff} &\stackrel{\text{def}}{=}
			\varphi_+ + \pi\left[1-\mathrm{sgn}\left(\cos{\varphi_-}\right)\right]	
		\end{split}
	\end{equation}
	where $\alpha_\mathrm{eff}$ acts as an effective, in-situ 
	tu{n}able 
	$\alpha$ of a SNAIL controlled by the flux bias parameter $\varphi_-$
	{and with a maximum value $\alpha_0$},
	while $\varphi_\mathrm{eff}$ corresponds to the SNAIL effective flux bias\cite{frattini_3-wave_2017},{controlled by $\varphi_+$ parameter}.
	Making use of the definitions \eqref{eff_quantities}, 
	{G-SNAIL bias parameters can be expressed as $b_\mathrm{e} = \alpha_\mathrm{eff}\cos{\varphi_\mathrm{eff}}$ and $b_\mathrm{o}=\alpha_\mathrm{eff}\sin{\varphi_\mathrm{eff}}$. Consequently,}
	G-SNAIL potential energy \eqref{potential} 
	{can be re-expressed as}
	\begin{equation}
		\label{effective_potential}
		\frac{U(\bm{\varphi})}{E_\mathrm{J}} =-3\cos{\frac{\bm{\varphi}}{3}}-\alpha_\mathrm{eff}\cos \left(\bm{\varphi} - \varphi_\mathrm{eff}\right)
	\end{equation}
	{which} is formally equivalent to the potential energy of a SNAIL \cite{frattini_3-wave_2017}.
	This one-to-one correspondence clarifies how a G-SNAIL behaves as a SNAIL with a tunable $\alpha$, thus can be employed in every SNAIL-based application for increased versatility. 
	{The bias parameters $b_\mathrm{e}$ and $b_\mathrm{o}$ can be defined for a SNAIL as well, where $\alpha_\mathrm{eff} = \alpha_0 = \alpha$
		and $\varphi_\mathrm{eff} = 2\pi\Phi/\Phi_0$, where $\Phi$ is the flux threaded to the loop in the right image of Fig. 1\textbf{a}.}
	As represented in Fig. 1\textbf{b}, the two bias knobs of the G-SNAIL allow to choose values for $b_\mathrm{e}$ and $b_\mathrm{o}$ on a 2D space thanks to the tunable $\cos{\varphi_-}$ term
	{in the $\alpha_\mathrm{eff}$, which enables}
	an independent tuning of even and odd order Josephson nonlinearities in the potential energy \eqref{potential}. Instead, for a SNAIL with an $\alpha$ set by fabrication, the
	{parameters}
	are constrained to live on the circumference of the circle in Fig. 1\textbf{b} and cannot be independently chosen. This peculiarity of our device provides additional flexibility for the tuning of expansion coefficients $c_n$ in expression \eqref{dipole_potential}.
	In a realistic G-SNAIL, where the Josephson energies of the small JTJs in the
	{right}
	{image} of Fig. 1\textbf{a} are never perfectly matched, $\alpha_\mathrm{eff}$ cannot be tuned all the way down to zero. This effect is not detrimental as usually SNAILs are operated at values of $\alpha$ not too close to zero.
	
	{When a G-SNAIL is configured with a symmetrical layout as in the right image of Fig. 1\textbf{a}, the phase biases $\varphi_{\pm}$ are proportional to $i_\mathrm{\pm} = i_1\pm i_2$, where $i_{1(2)}$ are the bias currents in Fig. 1\textbf{c}. The choice of a symmetrical layout then simplifies the control of $\alpha_\mathrm{eff}$ and $\varphi_\mathrm{eff}$, which can be independently set by $i_\mathrm{\pm}$, as can be obtained from equations \ref{eff_quantities}.}
	
	To demonstrate the capability of G-SNAILs to suppress the Kerr constant in a three-wave mixing parametric amplifier, we build an array of 20 G-SNAILs embedded in a $\lambda/2$ microstrip resonator (Fig. 1\textbf{c}) which is strongly coupled to a probe port (left) and weakly coupled to a pump port (right). The resonator would have a bare resonance frequency $\omega_\mathrm{r}{/2\pi} = 15.61 \rm{GHz}$, when substituting the array of G-SNAILs with a short circuit, and {a characteristic} impedance $Z_0 = 50\Omega$.
	The value for $\omega_\mathrm{r}$ was extracted from a resonance frequency fit as a function of bias phases $\varphi_{\pm}$, as explained below.
	{These quantities are obtained considering a lumped-element model of the resonator in Fig. 1\textbf{c}, where the array of G-SNAILs is in series with the equivalent linear inductance of the resonator $L_\mathrm{r} = Z_0/\omega_\mathrm{r}$.}
	The experimental device is fabricated with an aluminum-based single-step deposition, where the Al/AlOx/Al JTJs are realized with a Dolan bridge technique. {The bottom layer is deposited with a nominal thickness of 35nm, while the top layer is deposited with a nominal thickness of 120nm.}
	{From fitting resonance frequency data with the model discussed below, we extract a critical current for the big JTJs in the central array of the G-SNAILs $I_\mathrm{c}\approx12\mathrm{uA}$.}
	The top panel of Fig. 1\textbf{d} shows a Scanning Electron Microscopy (SEM) image of the center of the $\lambda/2$ resonator, with the embedded array of 20 G-SNAILs and the two flux lines, one for each side of the array. The bottom panel of the same figure shows a magnified picture of a portion of the array, where each G-SNAIL is made by an array of three big JTJs (purple) in the center shunted by two small JTJs (blue) on each side.
	\begin{figure}
		\label{fig2}
		\includegraphics[width = \columnwidth]{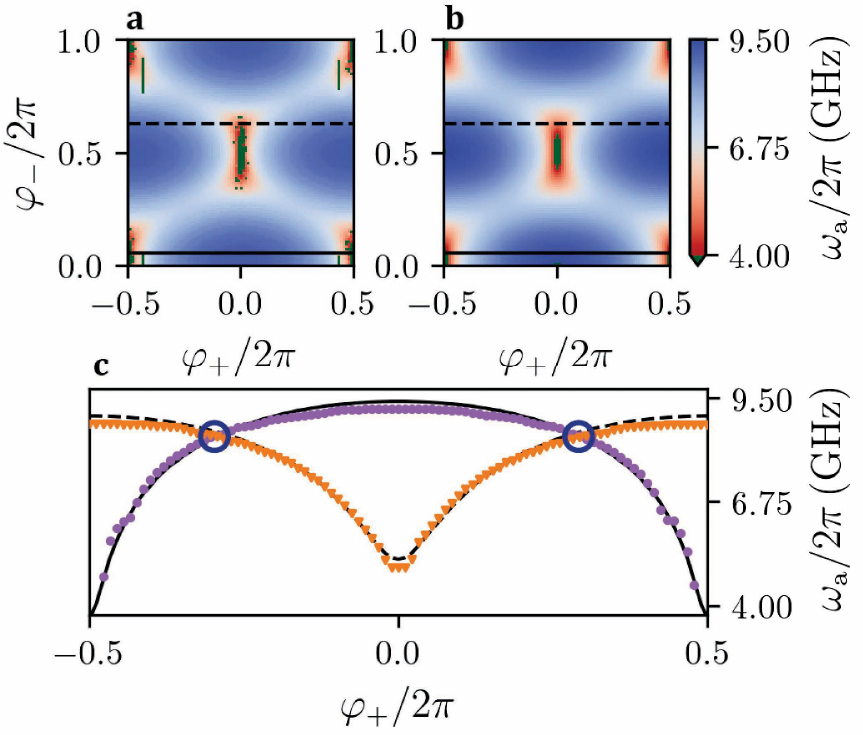}
		\caption{(a) Measured resonance frequency $\omega_\mathrm{a}$ at signal port of the device, while sweeping bias phases $\varphi_+$ and $\varphi_-$. Dark green color indicates below our experimentally measurable bandwdith, restricted by 4-12 GHz circulator. (b) Theory plot of $\omega_\mathrm{a}$ based on expression \eqref{omega} evaluated for a set of fit parameters: $\alpha = 0.145$, $x_\mathrm{J} = 0.053$ and $\omega_\mathrm{r}/2\pi = 15.61 \rm{GHz}$. (c) Comparison between measured and theory resonance data as a function of $\varphi_+/2\pi$, evaluated for two values of $\varphi_-$ corresponding to the two cuts in the 2D frequency landscapes. The two curves intersect in two specular points, highlighted with blue circles. One point corresponds, for each curve, to a same value of $\omega_\mathrm{a}$ but different combinations of $c_3$ and $c_4$ coefficients.}
	\end{figure}	
	
	As a first demonstration of the doubly-flux bias capability of our device, we perform a small{-}signal 1-tone spectroscopy at signal port with a vector network analyzer for many combinations of the bias phases $\varphi_{\pm}$. Then, we extract the resonance frequency with a tool for the analysis of microwave resonators \cite{probst_efficient_2015}.
	The resonance frequency predicted by theory reads
	\begin{equation} \label{omega}
		\omega_\mathrm{a}(\varphi_+,\varphi_-) = \frac{\omega_\mathrm{r}}{\sqrt{1+M\,{x_\mathrm{J}}/{c_2(\varphi_+,\varphi_-)}}}.
	\end{equation}
	where $\omega_\mathrm{r}$ is the bare resonator frequency, $x_\mathrm{J} = L_\mathrm{J}/L_\mathrm{r}$ is the ratio between the linear inductance $L_\mathrm{J} = \Phi_0/(2\pi I_\mathrm{c})$ of a big junction in the central array of the G-SNAIL and the equivalent inductance $L_\mathrm{r}$ of the bare resonator and $c_2$ is the second-order derivative of G-SNAIL potential energy \eqref{potential} evaluated at the value $\bm{\varphi}_0$ that minimizes the potential.
	Resonance frequency experimental data (Fig. 2\textbf{a}) acquired by varying $\varphi_{\pm}$ is compared to numerical simulations (Fig. 2\textbf{b}) based on expression (\ref{omega}). The simulations are performed using a set of parameters extracted with a nonlinear least-square fit of the experimental data. In Fig. 2\textbf{c} a comparison between measured (markers) and simulated (lines) resonance frequency is presented, for $\varphi_-/2\pi = 0.06$ (purple markers, solid lines) and for $\varphi_-/2\pi = 0.63$ (orange markers, dashed lines) while varying $\varphi_+$.
	From the comparison figure it is possible to notice how the two selected curves intersect in two specular points, highlighted with dark blue circles.
	This is a first demonstration of the advanced tuning capabilities of our device: at each intersection point in Fig. 2\textbf{c}, both curves have the same frequency but are associated to different combinations of $c_3$ and $c_4$ nonlinear expansion coefficients of the G-SNAIL potential energy \eqref{potential}. 
	
	We can now characterize the tunability of G-SNAILs nonlinear properties, in particular the Kerr coefficient $K$. For this sake, we measure the frequency Stark shift $\Delta_\mathrm{Stark}(\varphi_+,\varphi_-,\bar{n})$ as a function of the average number of photons populating the resonator $\bar{n}$ by fixing the bias phases and performing a 1-tone spectroscopy with a vector network analyzer while sweeping the probe tone power. The correspondence between probe tone power and $\bar{n}$ is calibrated from room temperature measurements of input line attenuation and resonance frequency and quality factors of the device\cite{probst_efficient_2015} extracted from small signal measurements.
	From this type of measurement, for each combination of $\varphi_{\pm}$, $K$ can be extracted as the frequency Stark shift per photon $\left.d\Delta_\mathrm{Stark}/d\bar{n}\right|_{\bar{n}=0}$. 
	\begin{figure}
		\label{fig3}
		\includegraphics[width = \columnwidth]{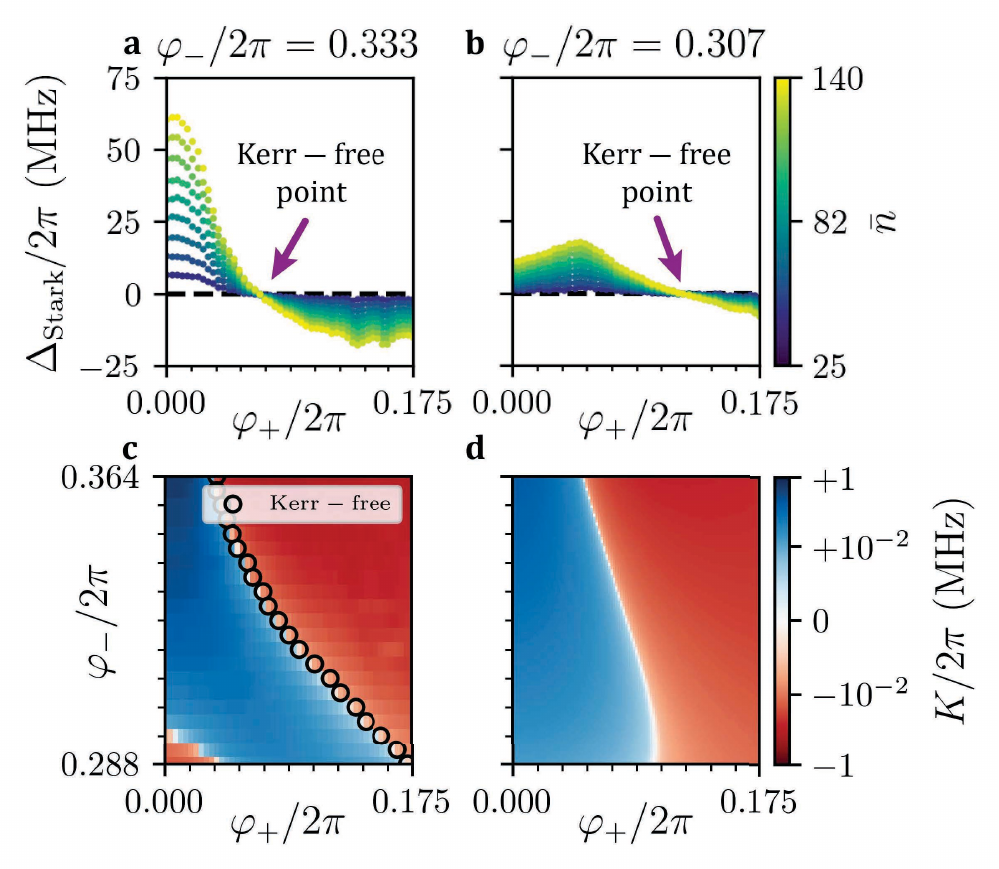}
		\caption{Measured Stark-shift data and Kerr.
			(a) Stark shift for a fixed value of $\varphi_-$. As a function of $\varphi_+$, the Stark shift can be positive, negative or zero. The shift is suppressed in correspondence of Kerr-free points, where resonance frequency does not depend on the average number of photons $\bar{n}$ populating the resonator.
			(b) Stark shift for a different value of $\varphi_-$ with respect to figure 3\textbf{a}. Here, the Kerr-free point corresponds to a different value of $\varphi_+$.
			(c) Measured Kerr constant as a function of $\varphi_{\pm}.$ Positive (blue) and negative (red) Kerr regions are connected by a \emph{Kerr-free line} line were $K=0$.
			(d) Theory plot of Kerr constant. A \emph{Kerr-free line} is predicted by theory, but it appears distorted with respect to the measured one.}		
	\end{figure}
	
	Stark shift experimental data is presented in Fig. 3\textbf{a}-3\textbf{b} for two fixed values of $\varphi_-$ over many values of $\varphi_+$. For each value of $\varphi_-$, the figures show positive, negative or zero Stark shift as a function of $\varphi_+$. Zero Stark shift points, where all the curves in Fig. 3\textbf{a}-3\textbf{b} intersect, can be identified as Kerr-free points \cite{frattini_optimizing_2018, sivak_kerr-free_2019}. 
	In Fig. 3\textbf{a}, for $\varphi_-$ corresponding to $\alpha_\mathrm{eff}=0.144$, the Kerr-free point appears at $\varphi_+/2\pi=0.07$ while, in Fig. 3\textbf{b}, for $\varphi_-$ corresponding to $\alpha_\mathrm{eff}=0.102$, the Kerr-free point appears at $\varphi_+/2\pi=0.123$.
	This is in accordance with the fact that, for a SNAIL, the flux value at which the Kerr-free point is located depends on the value of $\alpha$ \cite{frattini_optimizing_2018, sivak_kerr-free_2019}.
	From additional Stark shift measurements of the same type in Fig. 3\textbf{a}-3\textbf{b} we extract the Kerr coefficient as a function of $\varphi_+$ and $\varphi_-$, shown in Fig. 3\textbf{c}. In this figure it is possible to identify a \emph{Kerr-free line} (KFL), which marks the boundary between positive
	{(blue)}
	and negative
	{(red)}
	Kerr, and it corresponds to a continuous region of the $\varphi_\pm$ plane with suppressed $K$.
	This data demonstrates that it is indeed possible to keep $K=0$ for different combinations of bias phases, corresponding to different combinations of potential energy expansion coefficients $c_n$.
	In figure 3\textbf{d}, we plot the Kerr constant predicted by theory based on the model presented in [\onlinecite{frattini_optimizing_2018}]. The model is evaluated in the same range of $\varphi_\pm$ as in Fig. 3\textbf{c}, and for the same set of fit parameters as used for the theory plot of resonance frequency in Fig. 2\textbf{b}. The theoretically predicted KFL deviates from the measured one in figure 3\textbf{c}.
	{The formula used in the theoretical evaluation of the Kerr assumes that the G-SNAILs are all identical, and in each G-SNAIL the central array of three JTJs is made with identical elements. However, SEM inspection showed a 5\% variation of $\alpha$ along the array of G-SNAILs. Moreover, in each G-SNAIL, the array of three nominally indentical JTJs was not uniform, with the central JTJ having about 10\% bigger area than the other two. These imperfections could be responsible for the mismatch between measured and simulated Kerr.}
	
	We can now evaluate the span of resonance frequencies $\omega_\mathrm{KFL}$ over which the device can operate when flux biased along the KFL in Fig. 3\textbf{c}. We perform 1-tone spectroscopy measurements for many combinations of $\varphi_\pm$ selected along the KFL and extract the correspondent resonance frequencies. The measured data, shown in Fig. 4\textbf{a}, demonstrates that our device can operate with suppressed Kerr over a {continuous} range of frequencies of about 1.7 GHz.
	
	\begin{figure} \label{fig4}
		\includegraphics[width = \columnwidth]{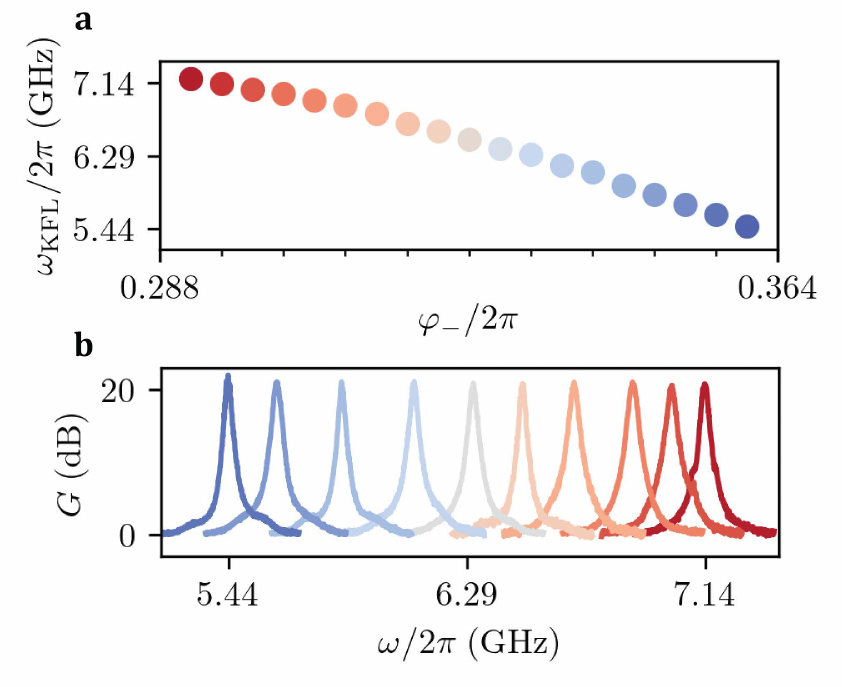}
		\caption{Device performances along a Kerr-free line. (a) When evaluated along a Kerr-free line, the resonance frequency $\omega_\mathrm{KFL}$ can be chosen on a range of 1.7 GHz. This proves that, with two flux-bias knobs, it is possible to suppress the Kerr {effect} while varying resonance frequency over a continuum. (b) Three-wave mixing gain over a Kerr-free line. By applying a resonant pump at frequency $\omega_\mathrm{p}=2\omega_\mathrm{KFL}$, this figure demonstrates that $c_3$ coefficient is non-zero all over the selected portion of the KFL, enabling three-wave mixing operations.}	
	\end{figure}
	
	Finally, to prove that our device is capable of three-wave mixing when flux biased along the KFL, we operate it as a parametric amplifier over the whole range of Kerr-free resonance frequencies in Fig. 4\textbf{a}. For each value of $\omega_\mathrm{KFL}$, we apply a pump at a frequency $2\omega_\mathrm{KFL}$ and measure the resulting gain curve while varying probe signal frequency $\omega$. Gain data, presented in Fig. 4\textbf{b}, show that it is possible to achieve 20 dB gain over the whole 1.7 GHz range in Fig. 4\textbf{a}. This demonstrates that for all the combinations of flux biases $\varphi_\pm$ belonging to the KFL in Fig. 3\textbf{c}, the expansion coefficient $c_3$ of potential energy \eqref{potential} can be kept large enough to ensure three-wave mixing operations.
	{This is in accordance with our theoretical analysis, which predicts $c_3$ variations across the experimentally measured KFL in Fig. 3\textbf{c} of less than a factor of 2 (not shown).}
	
	We want to clarify that the KFL in Fig. 3\textbf{c} does not exactly correspond to the region where we expect our device to have maximized dynamic range when operated as a parametric amplifier. In fact, the presence of a strong pump dresses the Kerr constant and shifts the positions of Kerr-free points from what predicted by Stark shift experiments \cite{sivak_kerr-free_2019}. Nonetheless, with an additional tuning of $\varphi_{+}$ and an {appropriate} detuning of pump frequency \cite{sivak_kerr-free_2019}, our device should be capable of delivering 20 dB parametric amplification with maximized dynamic range over a continuum of operating frequencies.
	This feature is currently under investigation in an optimized device
	{whose flux-lines embed low-pass filters to limit the amount of internal losses, which were appreciable in our device. Such losses are indeed detrimental for a correct characterization of the dynamic range, which can result overestimated by a variable amount depending on the flux-bias condition.}
	
	{In summary} in this Letter, we presented an experimental demonstration of the advantages that a doubly-flux biased G-SNAIL can provide in terms of potential energy engineering. In particular it was shown that, when employed as elementary blocks of a parametric amplifier, G-SNAILs provide enough tunability to the potential energy expansion coefficients $c_2$, $c_3$ and $c_4$ to suppress Kerr {effect} over a continuous region of the $\varphi_\pm$ plane, namely a Kerr-free line. Moreover, when flux biased along a KFL, our device was capable of three-wave mixing 20 dB amplification over a range of resonance frequencies of about 1.7 GHz.
	
	From our perspective, G-SNAILs are promising candidates as elementary blocks of a high performance three-wave mixing Traveling Wave Parametric Amplifier {(TWPA)}, a device that requires to simultaneously satisfy impedance- and phase-matching conditions. As the former is controlled by the linear response of the employed device, while the latter is strongly affected by Kerr \cite{zorin_josephson_2016}, we believe that the G-SNAIL could provide enough tunability to a TWPA in order to satisfy both requirements while being capable of three-wave mixing amplification, as suggested by the experimental data presented in this manuscript.
	Overall, the flexibility of G-SNAILs can be beneficial for all the applications based on superconducting circuits where it is required to properly balance the efficiency of two parametric processes to maximize performances \cite{ranadive_reversed_2021,zhang_engineering_2019}.
	\acknowledgments{A. Miano would like to thank D. Montemurro for useful discussions.
		This research was supported by the U.S. Army
		Research Office through Grant No.W911NF-18-1-0212. The view and conclusions contained in this document are those of the authors and should not be interpreted as representing the official policies, either expressed or implied, of the Army Research Office or the US Government. The US Government is authorized to reproduce and distribute reprints for Government purposes notwithstanding any copyright notation herein. Fabrication facilities use was supported by the Yale Institute for Nanoscience and Quantum Engineering (YINQE) and the Yale SEAS clean room. L.F. and M.H.D. are cofounders of, and L.F. is an equity shareholder in Quantum Circuits, Inc.}
	\bibliographystyle{ieeetr}
	\bibliography{bibliography}
\end{document}